\documentclass[prb,aps,twocolumn]{revtex4}
\usepackage{graphicx,amssymb,amsmath,color,psfrag}
\usepackage{amsthm}
\usepackage{amsfonts}
\usepackage{algorithmic}
\usepackage{enumerate}
\usepackage{latexsym}

\begin{document}
\newcommand{\dt}{\Delta\tau}
\newcommand{\al}{\alpha}
\newcommand{\ep}{\varepsilon}
\newcommand{\ave}[1]{\langle #1\rangle}
\newcommand{\have}[1]{\langle #1\rangle_{\{s\}}}
\newcommand{\bave}[1]{\big\langle #1\big\rangle}
\newcommand{\Bave}[1]{\Big\langle #1\Big\rangle}
\newcommand{\dave}[1]{\langle\langle #1\rangle\rangle}
\newcommand{\bigdave}[1]{\big\langle\big\langle #1\big\rangle\big\rangle}
\newcommand{\Bigdave}[1]{\Big\langle\Big\langle #1\Big\rangle\Big\rangle}
\newcommand{\braket}[2]{\langle #1|#2\rangle}
\newcommand{\up}{\uparrow}
\newcommand{\dn}{\downarrow}
\newcommand{\bb}{\mathsf{B}}
\newcommand{\ctr}{{\text{\Large${\mathcal T}r$}}}
\newcommand{\sctr}{{\mathcal{T}}\!r \,}
\newcommand{\btr}{\underset{\{s\}}{\text{\Large\rm Tr}}}
\newcommand{\lvec}[1]{\mathbf{#1}}
\newcommand{\gt}{\tilde{g}}
\newcommand{\ggt}{\tilde{G}}
\newcommand{\jpsj}{J.\ Phys.\ Soc.\ Japan\ }

\title{Spontaneous symmetry breakings in two-dimensional kagome lattice}
\author{Qin Liu $^{1,2}$}
\author{Hong Yao$^2$}
\author{Tianxing Ma$^{3,4}$}
\affiliation{$^1$ Institute of Microsystem and Information Technology, CAS, Shanghai
200050, China}
\affiliation{$^2$ Department of Physics, McCullough Building, Stanford University,
Stanford, CA 94305-4045}
\affiliation{$^3$Department of Physics and ITP, The Chinese University of Hong Kong, Hong
Kong}
\affiliation{$^4$Max-Planck-Institut f\"ur Physik Komplexer Systeme, N\"othnitzer Str.
38, 01187 Dresden, Germany}
\date{\today}

\begin{abstract}
We study spontaneous symmetry breakings for fermions (spinless and spinful)
on a two-dimensional kagome lattice with nearest-neighbor repulsive
interactions in weak coupling limit, and focus in particular on topological
Mott insulator instability. It is found that at $\frac{1}{3}$-filling where
there is a quadratic band crossing at $\Gamma$-point, in agreement with Ref.%
\cite{sun2009}, the instabilities are infinitesimal and topological phases
are dynamically generated. At $\frac{2}{3}$-filling where there are two
inequivalent Dirac points, the instabilities are finite, and no topological
phase is favored at this filling without breaking the lattice translational
symmetry. A ferromagnetic quantum anomalous Hall state with infinitesimal
instability is further proposed at half-filling of the bottom flat band.
\end{abstract}

\maketitle

%\author{Good Guys}

\textit{Introduction.} ---Recent interests have been revived in Fermi
surface instabilities in connection with dynamic generation of topological
Mott insulators \cite{sun2009,Raugh2008,Wu2004,Wu2007}. The first step is
forwarded by Wu and Zhang \cite{Wu2004,Wu2007} where a new mechanism of
generating spin-orbit coupling in strongly correlated, nonrelativistic
systems is proposed. Attentions are then focused in particular when the
Fermi surface shrinks into discrete points, usually as band-crossing points
(BCP), where the quasi-particle excitations are not described by the Fermi
liquid theory. Two examples are discussed in two-dimensional (2-D) honeycomb
\cite{Raugh2008} and checkerboard lattice \cite{sun2009} with $C_6$ and $C_4$
rotational symmetry respectively. In the honeycomb lattice \cite{Raugh2008},
the BCPs have linearly vanishing density of states (DOS) at the Dirac
points. Therefore only a full gap opening at the filling to the Dirac points
could facilitate gaining energy, and the nodal phase becomes unstable only
above finite critical interactions. While in the checkerboard lattice \cite%
{sun2009}, the BCP has quadratic dispersion with constant DOS in 2-D, so
that any gap opening at the BCP would help to gain energy and lead to the
\textit{infinitesimal} instabilities of the spontaneous breaking of
rotational or time-reversal symmetries (TRS).

\begin{figure}[tbp]
\begin{center}
\includegraphics[width=1.45in] {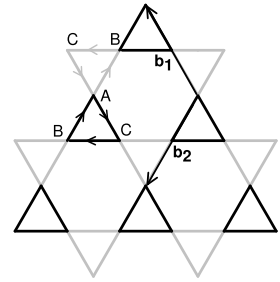}
\end{center}
\caption{Kagome lattice. There are three sites (``A, B, C'') per unit cell, $%
\mathbf{b}_1$, $\mathbf{b}_2$ are bravais vectors. The arrows on the links
represent currents respecting the reflection symmetry between up (dark) and
down (gray) triangles, by breaking which together with TRS, a quantum
anomalous Hall state is spontaneously generated.}
\label{kagome}
\end{figure}

Infinitesimal instabilities were proposed in 2-D systems with a quadratic
BCP, which is protected by TRS and $C_4$ or $C_6$ rotational symmetry \cite%
{sun2009,Sun2008}. In the explicit example with $C_4$ symmetry in
checkerboard lattice \cite{sun2009}, unequal next-nearest neighbor hoppings
connected and not by diagonal bonds are required to make the BCP quadratic,
which makes it hard to search for real materials. In this work, we suggest
to realize the infinitesimal topological instabilities in 2-D kagome lattice
with $C_6$ symmetry, where the kogom\'e compound herbertsmithite \cite%
{Helton2007,Jason2008} appears to be a good candidate. The kagome lattice is
a triangular lattice with three sites per unit cell, see Fig.\ref{kagome}.
Within only the nearest-neighbor (nn) hopping, it has three bands, the top
two bands cross at two inequivalent Dirac points and the bottom is a flat
band. However, interestingly, by considering a small next-nearest neighbor
(nnn) hopping, the flat band becomes dispersive, and its band structure
shows both quadratic and Dirac BCPs at $\frac{1}{3}$- and $\frac{2}{3}$%
-fillings separately, which facilitates us to stress the differences of
these two types of BCPs in a very same system. Although the magnetic
properties of kagome lattice were studied extensively \cite%
{Mielke1991,Bramwell2001,Castelnovo2008}, only little attention has been
paid to its nonmagnetic insulating behavior, especially the physics near the
quadratic BCP. Recently, topological phases with broken TRS and quantized
Hall conductance on kagome lattice have been studied for noninteracting
fermions \cite{Nagaosa2000,Guo2009,Yu2009}, while interacting fermions with
broken symmetries are rarely explored. Finally, we propose that a
ferromagnetic quantum anomalous Hall phase with infinitesimal instability
can be naturally realized at the half-filling of the bottom flat band.

\textit{Symmetries and order parameters.} ---Before discussing specific
models we first analyze the symmetries and define order parameters. The
symmetries of importance on kagome lattice under considerations are i)
reflection symmetry between up and down triangles, ii) $C_3$ rotational
symmetry around the centers of triangles, and iii) time-reversal symmetry.
We assume that the lattice translational symmetry wouldn't be broken,
namely, we do not consider the phases with finite $\mathbf{q=K_1}-\mathbf{K}%
_2$ in the susceptibilities, where $\mathbf{K}_1$ and $\mathbf{K}_2$ are the
two Dirac points. This is always true at $\frac{1}{3}$-filling because there
is only one BCP at $\Gamma$-point, and the instabilities occur only at $%
\mathbf{q}=0$ which preserves the lattice translational symmetry. Based on
the symmetry considerations, we introduce the order parameters as below.
Note that in the case of spinless fermions, since the order parameters,
defined only in the sublattice space with $\Psi^{\dagger}_k=(C^{%
\dagger}_{kA}C^{\dagger}_{kB} C^{\dagger}_{kC})$, have the same form as
those in the spinful case, we only write down the order parameters in
spinful case explicitly. For site order, the order parameters are $%
N^{\nu}(n^{\nu})=\langle\Psi|\tau^{\nu}\otimes\mathtt{I}_s(R_s)|\Psi\rangle$%
, where $\nu=0,1,2,3$ and $\Psi^{\dagger}_k=(C^{\dagger}_{kA\uparrow}C^{%
\dagger}_{kB\uparrow} C^{\dagger}_{kC\uparrow}C^{\dagger}_{kA\downarrow}
C^{\dagger}_{kB\downarrow}C^{\dagger}_{kC\downarrow})$. Here $\tau^{\nu}$
are Pauli matrices in spin space, and $\mathtt{I}_s=\text{diag}(1\; 1 \;1)$,
$R_s=\text{diag}(1\; e^{i\omega_0}\;e^{2i\omega_0})$ are 3 by 3 matrices in
sublattice space with $\omega_0=\frac{2\pi}{3}$. Among these order
parameters, those of interests are $n^0$ for nematic state (spontaneous
breaking of rotational symmetry) and $\vec{n}$ for nematic-spin-nematic
(NSN) state \cite{Wu2007,Kivelson2003}. This is because the charge ($N^0$)
and spin density wave ($\vec{N}$) order parameters remain constants in the
assumption of translational invariance, and in this assumption, the
ferromagnetic states double-meant by $\vec{N}$ is competitive only at
half-filling of the bottom flat band since we do not consider the spin-orbit
coupling here. For bond order, the order parameters are $\Delta^{%
\nu}_{u(d)}=\langle\Psi|\tau^{\nu}\otimes \mathtt{I}_b^{u(d)}|\Psi\rangle$
and $\delta^{\nu}_{u(d)}=\langle\Psi|\tau^{\nu}\otimes R_b^{u(d)}|\Psi\rangle
$, where the subscripts `u' and `d' indicate the order parameters defined in
up and down-triangles respectively, and
\begin{eqnarray}
&&\mathtt{I}_b^{u}=\left(%
\begin{array}{ccc}
0 & 1 & 0 \\
0 & 0 & 1 \\
1 & 0 & 0%
\end{array}%
\right),\;\mathtt{I}_b^{d}=\left(%
\begin{array}{ccc}
0 & 0 & e^{ix_1} \\
e^{-ix_2} & 0 & 0 \\
0 & e^{-i(x_1+x_2)} & 0%
\end{array}%
\right)  \notag \\
&&R_b^{u}=\left(%
\begin{array}{ccc}
0 & 1 & 0 \\
0 & 0 & e^{i\omega_0} \\
e^{2i\omega_0} & 0 & 0%
\end{array}%
\right) \\
&&R_b^{d}=\left(%
\begin{array}{ccc}
0 & 0 & e^{i(2\omega_0+x_1)}\nonumber \\
e^{-ix_2} & 0 & 0 \\
0 & e^{i(\omega_0-x_1-x_2)} & 0%
\end{array}%
\right)
\end{eqnarray}
with $x_i=\mathbf{k\cdot b}_i$, $i=1,2$, where $\mathbf{b_1}$ and $\mathbf{%
b_2}$ are bravais vectors of kagome lattice (see Fig.\ref{kagome}). The
order parameters under considerations are $\Delta^{\nu}_{u(d)}$ which don't
break the $C_3$ rotational symmetry for bond order.

\textit{Spinless fermions.} ---The model Hamiltonian for spinless fermions
with only nn repulsive interactions is
\begin{equation}
H=t\sum_{\langle
ij\rangle}(c^{\dagger}_ic_j+h.c.)-t^{\prime}\sum_{[ij]}(c^{%
\dagger}_ic_j+h.c.) +V\sum_{\langle ij\rangle}n_in_j  \label{Hamiless}
\end{equation}
where $t^{\prime}\cdot t>0$ is a small nnn hopping which is taken according
to the transfer integrals in a natural spin-1/2 kagome compound known as
herbertsmithite \cite{Jason2008,Helton2007}. The nnn hopping term makes the
bottom flat band dispersive and quadratic at the $\Gamma$-point with which
the Fermi surface crosses at $\frac{1}{3}$-filling. However this term won't
affect the low energy expansion near the Dirac points at $\frac{2}{3}$%
-filling. The nn repulsive interactions for side order will in general favor
the nematic phase, which breaks the $C_3$ rotational symmetry, with complex
order parameter $n=\langle n_A\rangle+e^{i\omega_0}\langle
n_B\rangle+e^{2i\omega_0}\langle n_C\rangle$, where $\langle
n_{\lambda}\rangle=\langle c^{\dagger}_{\lambda}c_{\lambda}\rangle$, $%
\lambda=A,B,C$. There are two independent order parameters in this phase,
the real part of $n$ depicts the picture that the charge densities are equal
on sites $B$ and $C$ but different on site $A$, while the imaginary part of $%
n$ visualizes that the charge densities on the three sites are all
different. At $\frac{1}{3}$-filling, the nematic order parameters split the
quadratic touching at $\Gamma$-point with $2\pi$ Berry phase \cite%
{Sun2008,Haldane2004} into two Dirac points with $\pi$ Berry phase each,
thus open a gap at $\Gamma$-point to gain energy. At $\frac{2}{3}$-filling,
the nematic order parameters pull the two Dirac points closer for small $V$.
While at large $V$, the two Dirac points with opposite Berry phases
annihilate each other and open a full gap to gain energy. Therefore the
nematic order is always a competing phase at both fillings. The phase
transition into nematic phase is of the first order. Our numerics show that
the nematic phase will stabilize at large $V$ for both fillings with $\text{%
Re}(n)=\pm 1$ respectively, where the charge densities reside all in $A$%
-site at $\frac{1}{3}$-filling and equally in $B$- and $C$-sites at $\frac{2%
}{3}$-filling.

For bond order, due to the frustration of the nn repulsive interactions, we
consider the possibilities of breaking reflection symmetry (or parity)
between up and down triangles and TRS, while conserve the $C_3$ rotational
symmetry ($\delta^{\nu}_{u(d)}=0$). The order parameters, $\Delta_{u(d)}$,
lead to four independent ones, $\Delta_{R\pm}=\frac{1}{2}[\text{Re}%
(\Delta_u)\pm\text{Re}(\Delta_d)]$ and $\Delta_{I\pm}=\frac{1}{2}[\text{Im}%
(\Delta_u)\pm\text{Im}(\Delta_d)]$. Among these, $\Delta_{R+}$ is only a
renormalization of the nn hopping energy $t$ and doesn't break any symmetry.
While $\Delta_{R-}$ breaks the parity and is responsible for the bond
density wave (BDW) phase where the hopping amplitude differs in up and down
triangles. This order parameter will remain the degeneracy of $\Gamma$-point
at $\frac{1}{3}$-filling, but opens a full gap at $\frac{2}{3}$-filling thus
gain energy. For the imaginary parts which break the TRS generally, $%
\Delta_{I+}$ respects the parity and corresponds to a staggered flux
picture, this order parameter won't open a gap at $\Gamma$-point,
and only shifts up-and-down the energies at the two Dirac points.
While $\Delta_{I-}$ breaks both parity and TRS giving rise to the
quantum anomalous Hall (QAH) phase with topologically protected edge
states \cite{Haldane1988} and opens a full gap at both fillings.

\begin{figure}[tbp]
\begin{center}
\includegraphics[width=2.4 in] {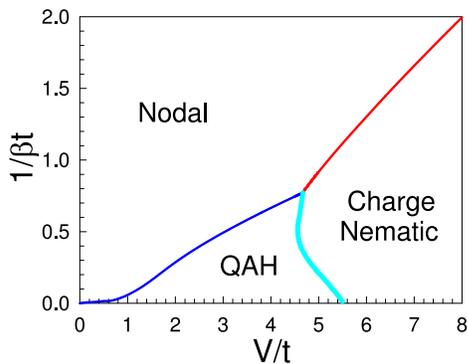}
\end{center}
\caption{(Color online). Finite temperature mean-field phase diagram for
spinless fermions in kagome lattice at $\frac{1}{3}$-filling with $t^{\prime
}/t=0.3$. Thick and thin lines are respectively second and first order
transitions.}
\label{phase}
\end{figure}

The mean-field Hamiltonian for spinless fermions is $H_{\text{MF}}=E_{0}+%
\frac{1}{L^{2}}\sum_{\mathbf{k}}\Psi _{\mathbf{k}}^{\dagger }[h_{0}(\mathbf{k%
})+h_{1}(\mathbf{k})]\Psi _{\mathbf{k}}$. The first term in the Hamiltonian
includes the nn and nnn hoppings, and
\begin{eqnarray}
&h_{1}&(\mathbf{k})=\nonumber \\
&-V&\left(\begin{array}{ccc} 2\langle n_{A}\rangle & \Delta
_{u}^{{\ast }}+\Delta _{d}e^{-ix_{2}} &
\Delta _{u}+\Delta _{d}^{{\ast }}e^{ix_{1}} \\
\Delta _{u}+\Delta _{d}^{{\ast }}e^{ix_{2}} & 2\langle n_{B}\rangle
&
\Delta _{u}^{{\ast }}+\Delta _{d}e^{-ix_{3}} \\
\Delta _{u}^{{\ast }}+\Delta _{d}e^{-ix_{1}} & \Delta _{u}+\Delta _{d}^{{%
\ast }}e^{ix_{3}} & 2\langle n_{C}\rangle
\end{array}
\right)\nonumber\\
\end{eqnarray}
with
$E_{0}=\frac{2V}{3}[\text{Re}(n)^{2}+\text{Im}(n)^{2}]+6V[\Delta
_{R+}^{2}+\Delta _{R-}^{2}+\Delta _{I+}^{2}+\Delta _{I-}^{2}]$. The
mean-field free energy is obtained as
\begin{eqnarray}
F(n,\Delta _{u},\Delta _{d})=E_{0}-\frac{1}{\beta L^{2}}\sum_{i\in
all}\log \left( 1+e^{-\beta E_{i}}\right)
\end{eqnarray}
where $\beta =1/k_{B}T$, and $L^{2}$ is the number of unit cells.
Since the order parameters are classified by different symmetries
they usually order at different $T_{c}$, therefore by minimizing the
free energy with respect to the order parameters separately, we
could study the mean-field phase diagram at finite temperatures. The
phase diagram at $\frac{1}{3}$-filling is shown in Fig. \ref{phase},
where we see that the instabilities are infinitesimal at this
filling, i.e., the symmetries are broken at low temperatures for
arbitrarily weak interactions. However this infinitesimal
instability is \textit{absent} in the phase diagram at
$\frac{2}{3}$-filling due to the vanishing DOS at Fermi level. Our
numerics show that at zero temperature, there are two quantum phase
transition points. For small $V$, the system remains in the
semi-metal phase, till $V_{c1}=1.85t$, the system
goes into the BDW phase, and finally stabilizes in the nematic phase for $%
V\geq V_{c2}=2.55t$. Note that the QAH phase never dominates in the phase
diagram at $\frac{2}{3}$-filling near the Dirac points without breaking the
lattice translational symmetry.

The infinitesimal instabilities are predicted in Ref.\cite{sun2009} in both
checkerboard and kagome lattices with $C_4$ and $C_6$ symmetries, where the
phase diagram is obtained in checkerboard lattice at half-filling. Here we
see that for spinless fermions on kagome lattice, similar phase diagram is
obtained at $\frac{1}{3}$-filling by lifting the degeneracies of the flat
band and make it quadratic at $\Gamma$-point. The infinitesimal instability
at $\frac{1}{3}$-filling on kagome lattice can be seen more clearly by
projecting the mean-field Hamiltonian into the two low energy states $%
\Phi^{\dagger}=(\phi^{\dagger}_1\; \phi^{\dagger}_2)$ near $\Gamma$-point as
\begin{eqnarray}
H^{\text{eff}}&=&\frac{1}{L^2}\sum_{\mathbf{k}}\Phi^{\dagger}_{\mathbf{k}}%
\left[h^{\text{eff}}_0-\frac{V}{2}(Q_1\sigma_z+Q_2\sigma_x+\Delta\sigma_y)%
\right]\Phi_{\mathbf{k}}  \notag \\
&+&\frac{V}{4}(Q^2_1+Q^2_2+\Delta^2)  \label{Hamileff}
\end{eqnarray}
where $h^{\text{eff}}_0=d_0\sigma_0+d_x\sigma_x+d_z\sigma_z$ with $%
d_0=(t-2t^{\prime})k^2$, $d_x=-2(t+t^{\prime})k_xk_y$, and $%
d_z=-(t+t^{\prime})(k_x^2-k^2_y)$, which have $d$-wave symmetry, and $%
Q_1=\langle\Phi^{\dagger}\sigma_z\Phi\rangle$, $Q_2=\langle\Phi^{\dagger}%
\sigma_x\Phi\rangle$ are the effective nematic order parameters, $%
\Delta=\langle\Phi^{\dagger}\sigma_y\Phi\rangle$ is the QAH order parameter.
Here $\sigma_i$'s are Pauli matrices in quasi-particle space. For weak
coupling, the ground state of Hamiltonian (\ref{Hamileff}) is QAH phase with
the gap function at $T=0$ given by $\Delta\propto\frac{1}{V}\exp[-\frac{1}{%
\rho_0V}]$, where $\rho_0$ is the constant DOS for a 2-D quadratic system.
We see that at the phase boundary the interactions have infinitesimal
instability, and the phase transition is of the second order. In contrast,
by projecting the original Hamiltonian into the two high energy states near
Dirac points, we have $d_0=t+2t^{\prime}$, $d_x=-\sqrt{3}tk_x$, and $d_z=%
\sqrt{3}tk_y$, which has $p$-wave symmetry. In this case, the gap function
for QAH phase is $\Delta\propto 1-\frac{\Lambda^2}{V^2}$ where $\Lambda$ is
the momentum cutoff, we see that at the phase boundary $V_c\simeq \Lambda$,
which is finite. We also perform the renormalization (RG) group analysis
based on the projected interacting model at both fillings. Specifically, the
result is very similar to that obtained in Ref.\cite{sun2009} at $\frac{1}{3}
$-filling. The interaction is marginal at the tree level, as expected from
the constant DOS at the quadratic BCP; while it is marginally relevant at
the one-loop level, which makes the system flow to strong coupling,
supporting our mean-field results for symmetry broken phases. Except for the
proportional constant which is model-dependent, the beta function is
qualitatively the same as that given in Ref.\cite{sun2009}. However at $%
\frac{2}{3}$-filling, the interaction is irrelevant at the one-loop level of
the RG and higher loops are quite involved.

\textit{Spinful fermions.} ---Now we take the spin degrees of freedom into
account and discuss the spin-$\frac{1}{2}$ fermions on kagome lattice. In
addition to the nn repulsive interactions, we also include an on-site
repulsive Hubbard term as well as a nn exchange term,
\begin{eqnarray}
H&=&t\sum_{\langle ij\rangle\sigma}(c^{\dagger}_{i\sigma}c_{j\sigma}+h.c.)
-t^{\prime}\sum_{[ij]\sigma}(c^{\dagger}_{i\sigma}c_{j\sigma}+h.c.)  \notag
\\
&+&U\sum_in_{i\uparrow}n_{i\downarrow}+V\sum_{\langle
ij\rangle}n_in_j+J\sum_{\langle ij\rangle}\vec{S}_i\cdot\vec{S}_j
\label{Hamilful}
\end{eqnarray}
where $n_i=n_{i\uparrow}+n_{i\downarrow}$. For site order, besides the
nematic order $n^0$ as discussed in the spinless case where there is no spin
order, there is a spin-triplet order parameter $\vec{n}$,
nematic-spin-nematic phase, where there is no charge order. Different from
the checkerboard lattice with $C_4$ symmetry, the spin degrees of freedom
won't restore the $C_6$ symmetry of kagome lattice in charge sector,
therefore we expect both nematic and NSN phases in the phase diagram. With
the spin degrees of freedom, the nematic order has contributions from both
the spin-singlet channel of Hubbard term $\frac{U}{4}n_in_i$, and the nn
repulsion term $Vn_in_j$, which together form an effective nn repulsion $%
V^{\prime}=V-\frac{U}{4}$ mimicing the nematic phase in the spinless model,
therefore the nematic phase exists only in the region $0<U<4V$ in the $U-V$
plane. For the NSN order, it comes only from the spin-triplet channel in the
Hubbard term $-\frac{U}{3}\mathbf{S}_i\cdot\mathbf{S}_i$ where $S^a_i=\frac{1%
}{2}c^{\dagger}_{i\alpha}\tau^a_{\alpha\beta}c_{i\beta}$. The NSN order
parameters shift the two degenerate quadratic touchings into four Dirac
points and open a gap at $\Gamma$-point thus gain energy. At $\frac{2}{3}$%
-filling, the NSN order parameters split the two Dirac points into four at
small $U$, one of the Kramer's pair will go closer and annihilate to open a
gap as $U$ increases, but the other two still remain linear touching and no
full gap is opened, so that not like the nematic order, the NSN order is not
competitive at $\frac{2}{3}$-filling.

The bond orders all come from the nn repulsive interactions $V$. The
spin-singlet ones $\Delta^0_{u(d)}$ are the order parameters of QAH phase as
discussed in the spinless model, while the spin-triplet ones $%
\vec\Delta_{u(d)}$ are the order parameters responsible for quantum spin
Hall (QSH) phase \cite{Raugh2008,Bernevig2006}. The QSH phase can be
visualized as double QAH layers with opposite flux, and in each layer, it is
a QAH picture which breaks simultaneously the parity and the TRS. There are
totally 12 independent spin-triplet order parameters which are $%
\Delta^{i}_{R\pm}= \frac{1}{2}\left[\text{Re}(\Delta^{i}_u)\pm\text{Re}%
(\Delta^{i}_d)\right]$ and $\Delta^{i}_{I\pm}= \frac{1}{2}\left[\text{Im}%
(\Delta^{i}_u)\pm\text{Im}(\Delta^{i}_d)\right]$ where $i=1,2,3$. Since the
full Hamiltonian (\ref{Hamilful}) has $SU(2)$ symmetry we could choose $%
\Delta^{\mu}_{u(d)}=(\Delta^0_{u(d)},0,0,\Delta^3_{u(d)})$ without loss of
generality, therefore the number of spin-triplet order parameters is reduced
into four. First, we note that $\Delta^3_{I-}$ and $\Delta^3_{R-}$ are
degenerate with $\Delta^0_{I-}$ and $\Delta^0_{R-}$ respectively, this is
because at the mean-field level, the effective hopping satisfies $%
t^{\uparrow}_{u(d)}=t^{\downarrow}_{d(u)}$ and $h_1^{\downarrow}=h_1^{%
\uparrow}(u\leftrightarrow d)$ which give the same spectrum due to the
lattice symmetry. Therefore the QAH and QSH phases gain equal energies in
the nn repulsive interactions $V$ at both fillings, and both $%
\Delta^{0,3}_{R-}$ are BDW order parameters. Second, the order parameter $%
\Delta^3_{R+}$ breaks only the $SU(2)$ symmetry and splits the energy bands
with different spins by shifting the energy at $\Gamma$-point up-and-down a
magnitude of $V|\Delta^3_{R+}|$ respectively, but keeps the quadratic point
touched. So the gap opened at $\Gamma$-point of magnitude $2V|\Delta^3_{R+}|$
is only due to spin-splitting without reducing the DOS. Finally, for the
exchange term $H_J=\frac{J}{4}[2n_i-n_in_j-2(c^{\dagger}_ic_j)(c^{%
\dagger}_ic_j)^{\dagger}]$, the $n_in_j$ term is only a shift of $V$ to $V-%
\frac{J}{4}$ and gains equal energy for both QAH and QSH phases. However,
the last term is the spin-singlet QAH order parameter, so that depending on
the sign of the exchange coupling, the QAH phase will be favored if $J>0$,
otherwise the QSH phase will dominate if $J<0$.

\begin{figure}[tbp]
\begin{center}
\includegraphics[width=4.5in] {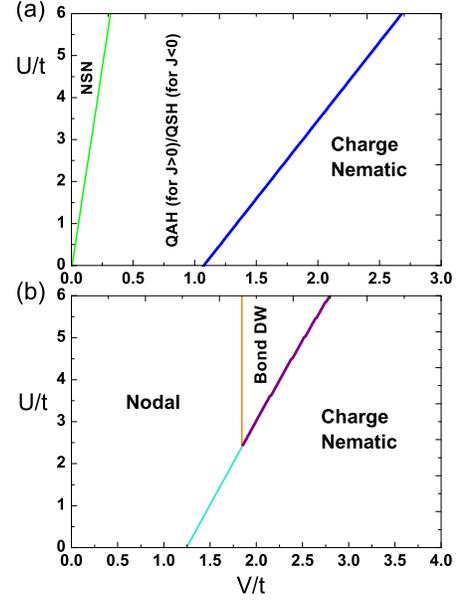}
\end{center}
\caption{(Color online). Zero temperature mean-field phase diagram for spin-$%
\frac{1}{2}$ fermions in kagome lattice at (a) $\frac{1}{3}$-filling (b) $%
\frac{2}{3}$-filling with $t^{\prime}/t=0.3$. Thick and thin lines are
respectively second and first order transitions. The phase diagram at $\frac{%
2}{3}$-filling with small $V$ and large $U$ should not be taken too
seriously.}
\label{phasespin}
\end{figure}

The mean-field free energy at zero temperature is $F=E_0+\frac{1}{L^2}%
\sum_{i=1}^{N_f}E_i$ where $E_0=\frac{2}{3}V^{\prime}|n^0|^2+\frac{U}{18}%
|n^3|^2+ 3V\sum_{\nu=0,3}\left(\Delta_{R+}^{\nu 2}+\Delta_{R-}^{\nu 2}
+\Delta_{I+}^{\nu 2}+\Delta_{I-}^{\nu 2}\right)$, and $N_f=6L^2f$ with the
filling $f=\frac{1}{3}$ and $\frac{2}{3}$. By minimizing the free energy,
the quantum mean-field phase diagrams in the $U-V$ plane are shown in Fig.%
\ref{phasespin}(a) at $\frac{1}{3}$-filling and in Fig.\ref{phasespin}(b) at
$\frac{2}{3}$-filling. At $\frac{1}{3}$-filling, it is seen again that all
the broken symmetries have infinitesimal instabilities. We first notice that
the spin-splitting phase (with the order parameter $\Delta^3_{R+}$) won't
win over any other competing phase in the entire $U-V$ plane, which agrees
with our analysis above. When $0<4V<U$ where the effective interaction for
nematic order $V^{\prime}$ is attractive, only the NSN and the QAH/QSH
orders compete. Starting from the $U=0$ axis, the topological phases win
first for arbitrarily weak nn repulsion $V$, and then be suppressed by the
NSN phase at large $U$. When $0<U<4V$ where all nematic, NSN and topological
phases compete, the NSN order is completely suppressed. The topological
phase dominates at small $V$ but then be taken over by the nematic phase, as
seen in the spinless case, and the large on-site $U$ helps to reduce the
nematic order further. At $\frac{2}{3}$-filling, all the instabilities are
finite again, and the topological phases are not favored at all. Compared
with the phase diagram in spinless case, we notice that at $U=0$, the
multi-channel repulsions $\uparrow\uparrow$, $\downarrow\downarrow$, $%
\uparrow\downarrow$ and $\downarrow\uparrow$ in the nn repulsion $V$ in
spinful model, facilitate the ordering of nematic phase relative to BDW
phase, therefore the BDW phase disappears in the $U=0$ axis, and the nematic
phase begins to order at a finite interaction $V>V_c=1.25t$. However, this
multi-channel advantage is suppressed by large on-site Hubbard term, and the
BDW appears for $U\geq U_c=2.44t$. In the phase diagram at $\frac{2}{3}$%
-filling, we need to point out that the nodal phase at small-$V$ and large-$U
$ should not be taken too seriously. Since there are two Dirac points at
this filling, the system could in principle gain energy by breaking the
lattice translational symmetry and nesting between them, which is the case
we do not consider in our discussions.

Finally we discuss in brief the idea of ferromagnetic QAH state with
infinitesimal instabilities at half-filling of the bottom band on kagome
lattice. If we consider only the nn hopping $t$, the bottom band is
completely flat. In the presence of an on-site repulsive Hubbard term, its
ground state is ferromagnetic at $\frac{1}{6}$-filling \cite{Mielke1991}.
Now if we turn on the nn repulsion $V$, where the physics is effectively
depicted by the spinless model at $\frac{1}{3}$-filling discussed before,
the QAH state with infinitesimal instabilities will be favored in the
ferromagnetic background. The ferromagnetic QAH phase is a topological state
with quantized Hall conductivity $\sigma_{xy}=\frac{e^2}{h}$, it is stable
even in the presence of a small nnn hopping $t^{\prime}$. The ferromagnetic
QAH state has been proposed in Hg$_{1-y}$Mn$_y$Te quantum wells with
Mn-doped impurities \cite{lcx2008}, where a small magnetic filed is still
needed to polarize the Mn moments. However, on kagome lattice, we only need
to tune the filling to right to realize the ferromagnetic QAH state. The
detail study of the ferromagnetic QAH state on kagome lattice will be
present in a separate work \cite{Yao2009}.

The material ZnCu$_3$(OH)$_6$Cl$_2$, known as herbertsmithite, appears to be
an excellent realization of the 2-D spin-$\frac{1}{2}$ kagome lattice \cite%
{Jason2008,Helton2007}. This material consists of Cu kagome layers separated
by nonmagnetic Zn layers, structurally with space group $R\bar 3 m$ and
lattice parameters $c=14.049\mathring{A}$ which is twice of $a=b=6.832%
\mathring{A}$. The transfer integrals $t$ and $-t^{\prime}$ of this material
are $87$meV and -$10$meV respectively \cite{Jason2008}, which implies the
opposite sign between nn and nnn hoppings and their ratio is quantitatively
in agreement with that taken in our numerics. We suggest to realize all the
symmetry broken phases on spin-$\frac{1}{2}$ kagome lattice by tuning the
filling, say replacing the Cl$^-$ ions with sulfur (S) or oxygen (O), in
this kind of material.

\textit{Acknowledgements.} ---Part of the present work was done when
the first author was at Department of Physics, Fudan University. The
authors would like to thank S. Raghu and X.-L. Qi for helpful
discussions. Q.L. acknowledges the support of China Scholarship
Council for support. Tianxing Ma acknowledges the support of CUHK
Direct Grant 2060374.

\end{document}